\documentclass[letterpaper]{natureprintstyle}
\usepackage{graphicx,epsfig}
\usepackage{amsmath,amssymb,bbm}
\usepackage{bm}
\usepackage{color}
\usepackage[usenames,dvipsnames]{xcolor}
\usepackage[colorlinks=true,linkcolor=Red,citecolor=Green,linktoc=page]{hyperref}
\usepackage{multirow}

\usepackage[varg]{txfonts}
\usepackage{ulem}
\usepackage{fancyhdr}
\usepackage[caption=false]{subfig}
\usepackage{balance}
\usepackage{float}
\usepackage{flushend}
\usepackage{graphicx}
\usepackage{amsmath}

\DeclareFontFamily{U}{rcjhbltx}{}
\DeclareFontShape{U}{rcjhbltx}{m}{n}{<->rcjhbltx}{}
\DeclareSymbolFont{hebrewletters}{U}{rcjhbltx}{m}{n}

\usepackage[caption=false]{subfig}
\DeclareCaptionListOfFormat{nparens}{#1#2}
\usepackage[varg]{txfonts}
\usepackage{ulem}
\usepackage{fancyhdr}
\newcommand{\addresses}[1]{
\thispagestyle{fancy} \lfoot{\parbox{\textwidth}{ \vspace{0.3cm}
 \rule{\textwidth}{0.2pt}
\hspace{-0.2cm} \textsf{\scalefont{0.80} #1} \vspace{-0.2cm}
\begin{center}{\scalefont{0.87} \thepage}\end{center}}}
\cfoot{} }

\DeclareMathSymbol{\lamed}{\mathord}{hebrewletters}{108}

\begin{document}
\title{Interference, diffraction, and diode effects in superconducting array based on Bi$_{0.8}$Sb$_{1.2}$Te$_3$ topological insulator}

\author{
	Xiangyu \,Song,$^{1}$
	Soorya Suresh-Babu,$^{1}$
    Yang \,Bai,$^{1}$
	Dmitry \,Golubev,$^{2}$
	Irina \,Burkova,$^{1}$
	Alexander \,Romanov,$^{1}$
	Eduard \,Ilin,$^{1}$
	James \,N.\,Eckstein$^{1}$
	and Alexey \,Bezryadin$^{1}$
}


\maketitle
\addresses{
$^{1}$ Department of Physics, University of Illinois at Urbana-Champaign, Urbana, IL 61801, USA;
$^{2}$ Pico Group, QTF Centre of Excellence, Department of Applied Physics, Aalto University School of Science,
Aalto 00076, Finland}

{\bf
It is a well known phenomenon in optics that spectroscopic resolution of a diffraction grating is much better compared to an interference device having just two slits, as in the Young's famous double-slit experiment. On the other hand, it is well known that a classical superconducting quantum interference device (SQUID) is analogous to the optical double-slit experiment. 
Here we report experiments and present a model describing a superconducting analogue to the diffraction grating, namely an array of superconducting islands positioned on a topological insulator (TI) film Bi$_{0.8}$Sb$_{1.2}$Te$_3$. In the limit of extremely weak field, of the order of one vortex per the entire array, such devices exhibit a critical current peak that is much sharper than the analogous peak of an ordinary SQUID. Because of this, such arrays can be used as sensitive absolute magnetic field sensors. An important finding is that, due to the inherent asymmetry of such arrays, the device also acts as a superconducting diode.}

\bigskip

{\bf Introduction} 

  Topological insulator (TI) films can be made superconducting, using the the proximity effect, by placing superconducting electrodes on the TI surface\cite{Eckstein1}. Such hybrid structures provide a testing ground for topological superconductivity\cite{Frolov}. They continue to be a hot topic in condensed matter physics due to the predicted and, to some extent, observed signatures of Majorana zero modes\cite{VanHarlingen1}. Although single junctions and SQUIDs have been previously studied in great depth, topological superconductor arrays and superconducting quantum interference filters (SQIFs) have not been sufficiently investigated. Various SQUIF systems\cite{Sch1,Sch2} are interesting for applications, since they contain many interfering superconducting loops and thus enable absolute magnetic field sensitivity\cite{Sch3}. Moreover, arrays can provide room for multiple interacting vortices, which may be subjected to quantum braiding manipulations. These are sought after due to the promise of topologically protected quantum computation.
  
  Another phenomenon which has attracted attention recently is the superconducting diode effect\cite{Ideue,Souto,Chen,Wakatsuki,Baumgartner}. A magnetically controllable superconducting diode has been demonstrated in an artificial superlattice [Nb/V/Ta]$_\mathrm{n}$ without a center of inversion\cite{Ando}. Previously, superconducting rectifiers have been realized in asymmetric superconducting nanowire loops\cite{Gurtovoi, Murphy}. Generally speaking, nonreciprocal phenomena are well known in relation to semiconductor diodes with a p–n junction. They exhibit either a high or a low resistance, depending on the current polarity. The diode effect is used in a number of very important electronic components, including photodetectors, ac rectifiers, and frequency multipliers. But, due to their finite resistance, Joule heating and energy losses are inevitable in such devices. Therefore, a superconducting rectifier or a diode, characterised by zero resistance, remains highly desired for computational, sensing, and communication applications with ultralow power consumption. Such device should have a zero resistance at one current polarity and a nonzero resistance at the opposite polarity. It was recently reported that Mo-Ge perforated films can be patterned using a conformal mapping approach in order to create superconducting diodes\cite{Kwok}. 
  
  The focus of our study is a square array of superconducting Niobium islands overlaying an intrinsic Bi$_{0.8}$Sb$_{1.2}$Te$_3$ topological insulator film (Nb-BST-Nb array). The system acts as the superconducting analogue of a diffraction grating. We observe an extreme sensitivity of the array to the external magnetic fields because of the diffraction-grating-style interference the condensate flows through in all parallel junctions. The interference peak is so narrow that it allowed us to distinguish magnetic fields which differ by about 8 nT, using a simple dc technique. The zeroth-order critical current interference peak is significantly sharper and taller than all other maxima, thus allowing absolute magnetic field measurements. By contrast, regular superconducting quantum interference devices\cite{Tinkham, Barone} are characterized by a periodic dependence on the external field and thus do not allow absolute field measurements. We also observe a strong Fraunhofer diffraction effect, which leads to a complete suppression of the second interference peak. This fact, namely the complete suppression of the supercurrent, provides a testing ground for the search of Majorana fermions. In the ideal case, when the flux per junction equals one flux quantum, the critical current is zero unless Majorana states are present, as they have a different current-phase relationship\cite{VanHarlingen1}.
  
  Our experiments demonstrate that topological insulator-superconductor arrays can behave as efficient superconducting rectifiers. In other words, the system of proximity-coupled islands acts as a superconducting diode, i.e., it exhibits a dependence of the critical current on the polarity of the bias current. Our design might be superior in some aspects with respect to previously published designs. For example, our system requires only very small magnetic fields, of the order 10 $\mu$T, for the occurrence of the diode effect. Such fields are orders of magnitude lower than those used in a previously reported system\cite{Kwok}. To explain the results we propose a model which takes into account the high kinetic inductance of surface states of the topological insulator. Another key ingredient of the model is the naturally occurring variation in the Nb-BST junctions parameters. 

{\bf Experiment} 

 To fabricated superconducting arrays we use high quality topological insulator (TI) thin films (40 nm thick) of nominal composition Bi$_{0.8}$Sb$_{1.2}$Te$_3$ (BST) grown by molecular beam epitaxy.  Similar films were used in a study of the superconducting proximity effect in which it was shown that this composition provided undoped topological insulators without bulk carriers or an accumulation layer caused by band bending. The only free carriers in this material are the topological surface carriers due to the inter-band Dirac cone that exhibits spin-momentum locking.  There it was shown that the proximity effect was restricted to the surface in contact with the superconductor.\cite{Hlevyack}

\begin{figure}[t]
\centering
\includegraphics[width=8cm]{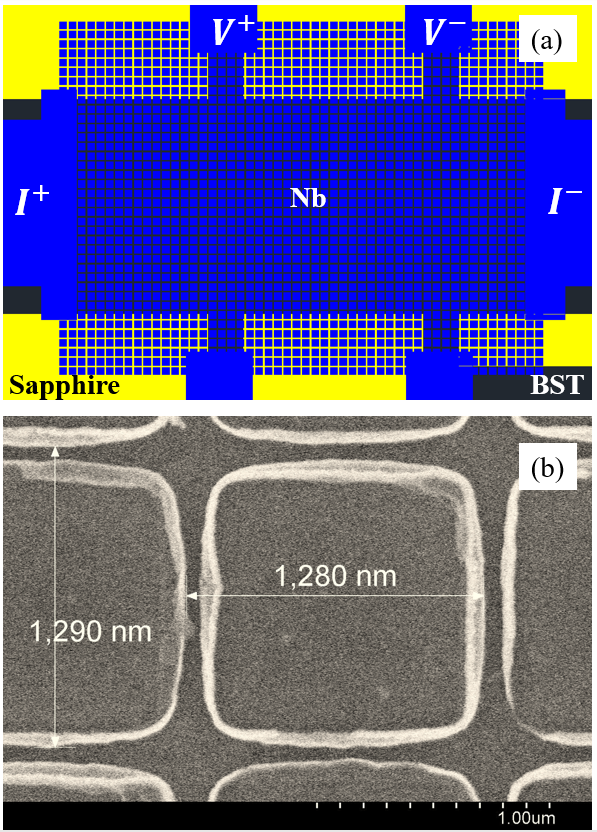}
\caption{ (a) Autocad-generated design file of the device. A square array of square Niobium islands is placed over the surface of a topological insulator epitaxial film, Bi$_{0.8}$Sb$_{1.2}$Te$_3$ (BST). The blue color illustrates the Nb leads and the square Nb islands. The black color represents the BST film. The yellow color is where the BST film was milled off to expose the sapphire substrate underneath, thus create trenches to isolate devices from each other. The bias current is applied horizontally between the $I^{+}$ and $I^{-}$ leads and the magnetic field is oriented perpendicular to the surface of the sample. The voltage is probed on the $V^{+}$ and $V^{-}$ contact pads. (b) An SEM image of the Nb array.}
\label{SEM} 
\end{figure}

Our Nb-BST-Nb arrays consist of Niobium
square islands arranged into a square lattice deposited on top of the TI-BST film. The thickness of the islands is 30 nm and the lattice constant is $w+d=1.3 \mu$m (Fig.\ref{SEM}). The array is fabricated by electron beam lithography and plasma sputtering of the Nb film and contains 23 x 23 Nb islands (see Methods for further details). The gap between neighboring islands, as estimated from the SEM images, equals $d\approx$150 nm (Fig.\ref{SEM}(b)). So, the width of each island is approximately $w\approx1.15\mu$m.

All experiments were performed at a temperature $T=$ 0.3K. The voltage and the current leads were carefully filtered against electromagnetic noise and well thermalized. The filtering was done by winding a highly resistive twisted-pair of constantan wires around a copper cold finger and coating them with stycast epoxy filled with Cu particles, which efficiently absorbs parasitic electromagnetic noise. For the purpose of thermalization, varnish-coated Cu wires have been squeezed between Cu bars maintained at 0.3K. All electrical signals were passed through such cold Cu wire segments to ensure that all leads attain the base temperature. In addition to the above measures, 1.9MHz $\pi$-filters were installed at the top of the cryostat. Also, a 100 kOhm resistor was placed right at the BNC input leading to the current biasing connection of the sample. This resistor served to further reduce the interference of the external noise. The high efficiency of our filters was confirmed by the fact that the observed fluctuations of the critical current were extremely low, only about 0.1$\%$. 

\begin{figure*}[t!]
\centering
\includegraphics[width=16cm]{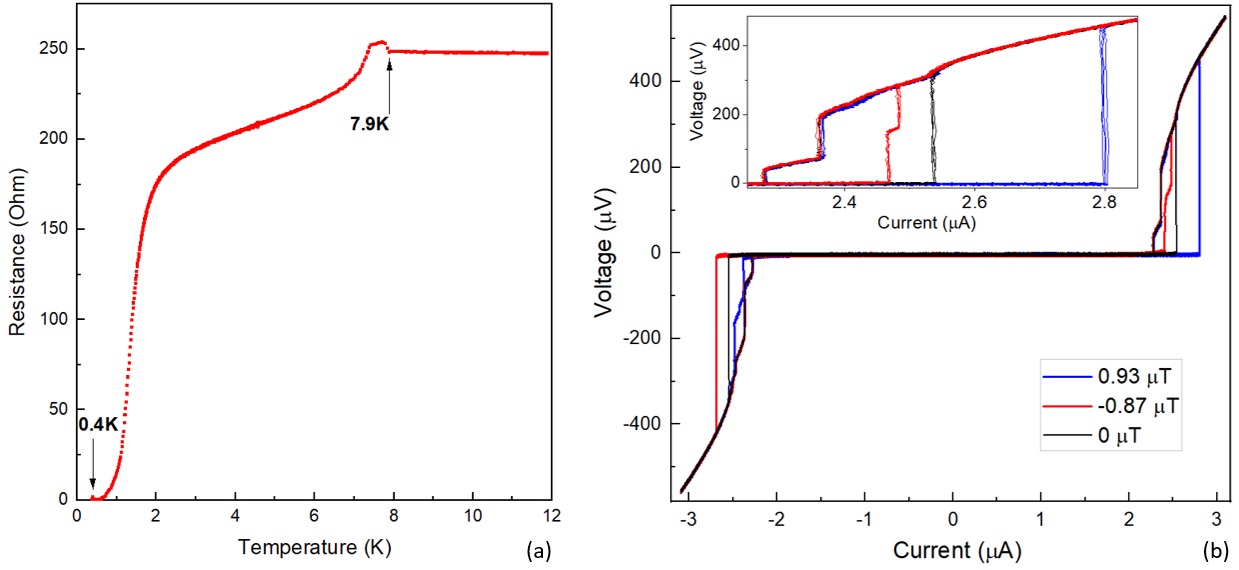}
\caption{(a) Resistance versus temperature (R-T) curve. (b) Examples of voltage-current (V-I) curves, taken at different (weak) magnetic field values and at $T=$ 0.3K. The V-I curve measured at zero field (black) is presumed symmetric, while the others are not. The curves illustrate the strong sensitivity of the array to the magnetic field. (b, Insert) A zoomed in version of Fig.2b. Each curve is measured four times. The curves illustrate the fact that the switching current does not change from one measurement to the next one if the magnetic field is fixed. }
\label{Fig_RT-VI}
\end{figure*}

The resistance versus temperature (R-T) curve (Fig.\ref{Fig_RT-VI} (a)) shows that a superconducting transition takes place in the Nb islands first, at $T\approx$ 8K. Due to the proximity effect, the surface of the topological insulator becomes superconducting near the islands. Consequently, the resistance drops to zero at $T\approx$ 0.4 K. Global superconductivity of the array is evident from the voltage-current (V-I) curves (Fig.\ref{Fig_RT-VI} (a)), which show a critical current below which the voltage is zero. So, voltage-current (V-I) curves exhibit sharply defined critical currents (Fig.\ref{Fig_RT-VI}(b)). Note that in order to obtain such V-I curves it was necessary to apply a small magnetic field to compensate for the Earth's field. The symmetric curve (black) is presumed to occur at zero magnetic field (Fig.\ref{Fig_RT-VI}(b)). A zoomed-in version of these three V-I curves is shown in Fig.\ref{Fig_RT-VI}(b, Insert). The V-I curve was measured four times at each magnetic field. Such repeated measurements demonstrate the same switching current. The fluctuations are less than 0.1$\%$, indicating, among other things, that noise, thermal phase slips, as well as quantum tunneling of phase slips do not contribute significantly to cause premature switching events. 

\begin{figure}[t]
\centering
\includegraphics[width=8cm]{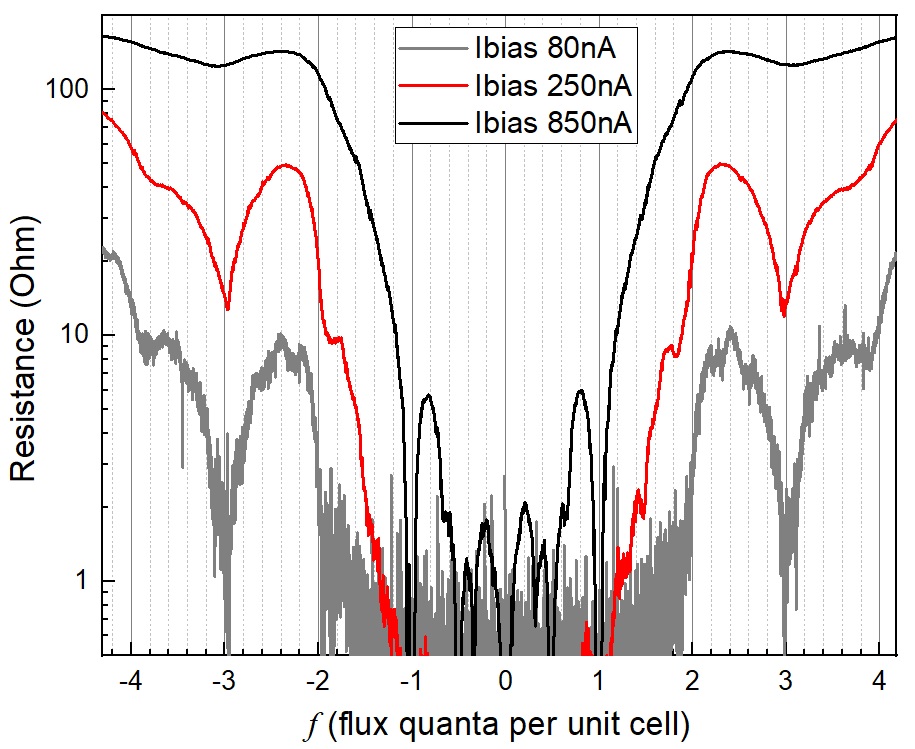}
\caption{Resistance versus magnetic field (R-B) curves. Three curves were taken at different amplitudes of the ac bias current amplitudes. The temperature was $T=$ 320mK.}
\label{RB}
\end{figure}

\begin{figure}[t]
\centering
\includegraphics[width=9cm]{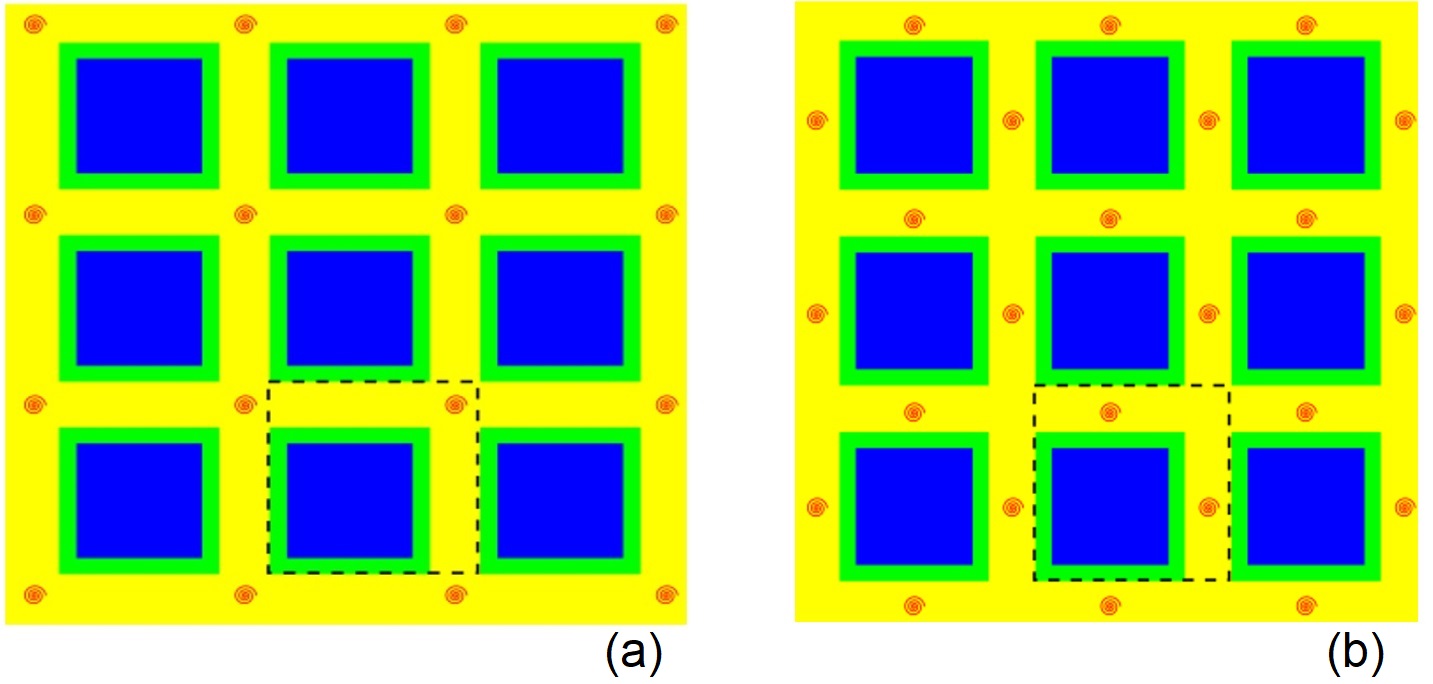}
\caption{a) Vortex cores distribution for the case $f=1$. A unit cell is shown by a dashed line. The Nb squares are shown as blue and the cores of vortices are shown, schematically, as swirls. The edges of the Nb squares are colored green to illustrate the regions when the Meissner effect is incomplete. The yellow color is the topological insulator. b) Vortex cores distribution for the case $f=2$.}
\label{f1vortices}
\end{figure}

\begin{figure}[t]
\centering
\includegraphics[width=8.5cm]{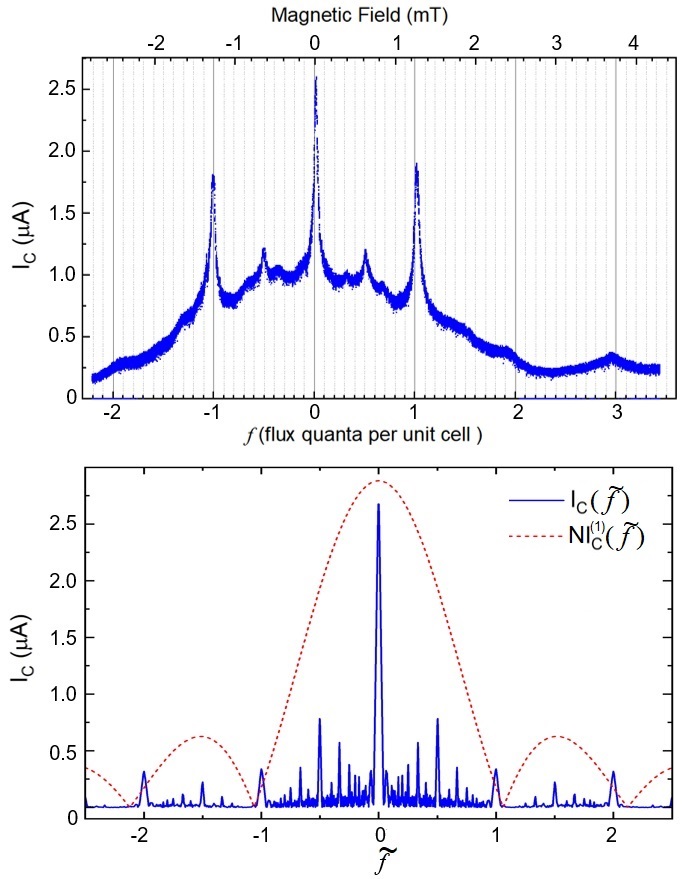} 
\caption{\label{Ic-B} \textbf{Critical current versus the normalized flux per unit cell, $f$.} 
Top panel: the critical current shows sharp peaks coinciding with integer values of $f$. The peak $f=2$ is suppressed.
Bottom panel, blue line: theoretical dependence $I_C^+(\tilde f)$ obtained by numerical solution of Eqs. (\ref{IJ},\ref{phi1}).
Bottom panel, red dashed line: the dependence $I_C^{(1)}(\tilde f)$ for a single junction (\ref{IC1}) multiplied by $N=23$.}
\label{Ic_B}
\end{figure}

\begin{figure}[!b]
\centering
\includegraphics[width=5.5cm]{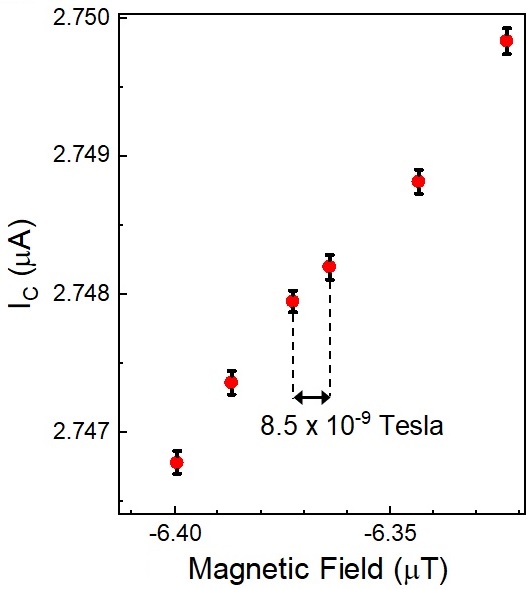}
\caption{\label{Sensitivity} Critical current sensitivity measurement. Each red dot represents the mean value of 1000 critical current measurements taken at 10Hz at a constant magnetic field on the left side of the $I^{-}$ central maximum, where the slope is the largest.
}
\label{Sensitivity}
\end{figure}

{\bf Results} 

A typical  resistance versus magnetic field (R-B) curve is shown in Fig.\ref{RB}. The horizontal axis represents the external magnetic field applied perpendicular to the sample plane. The field is normalized as
\begin{equation}
f=AB/\phi_0,
\label{f_def}
\end{equation}
where $A=(w+d)^2=1.69\mu$m$^2$ is the area of the unit cell of the square lattice, $B$ is the external magnetic field, and $\phi_0$ is the magnetic flux quantum. Thus, the oscillation period, defined as $f=1$, corresponds to the theoretical value $\Delta B=1.22$mT. The experimental value, $\Delta B=1.23$mT, was used when normalizing the field. The resistance is measured by calculating the slope of the best linear fit to the V-I curve. The V-I curve is measured using a low frequency ac signal, the amplitude of which is indicated on the graph. The V-I is nonlinear, thus explaining why the resistance is generally larger for larger amplitudes of the bias current (Ibias). 

The main features of the R-B plot are the presence of sharp minima in the resistance at $f=0$ and $f=1$ and a well pronounced minimum at $f=3$. However, the expected minimum at $f=2$ is missing. 

The existence of such sharp resistance minima can be understood as a result of vortices, which, in the range of the relatively weak fields considered here, are localized in the N-regions of the SNS junctions of the array. Here "S" stands for superconductor (Nb) and "N" represent the conducting normal surface states of the topological insulator. These states are, in fact, also weakly superconducting due to the proximity effect. Generally speaking, the resistance is defined by the level of dissipation in the array, which is proportional to the concentration of vortices and their mobility. Thus, the minimum at $f=0$ is explained by the lack of vortices in the array at zero magnetic field. On the other hand, the minima $f=1$ and $f=3$ correspond to cases where the number of vortices per unit cell is an integer (equal $f$). In the cases of $f$ being an integer, the vortex lattice is commensurate with the Nb islands lattice. Such commensurability ensures a pronounced minimum of the sample resistance, strong pinning of the vortices, and low mobility of the vortices which corresponds to a low dissipation. 

Recently, it has been demonstrated explicitly that vortices in SNS junctions have normal cores\cite{Roditchev}. We suggest that in the case of perfect commensurability, $f=1$, each vortex core is located between the corners of four neighbouring squares (Fig.\ref{f1vortices}(a)), where the proximity-induced superfluid density is the lowest. Thus, each vortex is located at the very minimum of its potential energy and so has a low mobility. Interestingly, a weaker but still prominent commensurability-related suppression of the resistance also occurs at fractional frustration values, namely $f=1/3, 1/2, 2/3$. This is a clear sign that vortices form an ordered lattice, even when they occupy every second or every third cell of the array while the other cells are free of vortices.

The missing minimum at $f=2$ (see the definition Eq.\ref{f_def}) can be explained by a qualitatively different arrangement of the vortices. With two flux quanta per unit cell ($f=2$), each junction receives one vortex. Such a conclusion stems from the fact that in the square array, there are exactly two junctions per unit cell (Fig.\ref{f1vortices}), the vertical one and the horizontal one. Thus, at $f=2$ the flux per junction equals one flux quantum. According to the Fraunhofer formula, the critical current of a single junction is $I_{c1}(f_1)=I_{c1}(0)\sin({\pi f_1})/(\pi f_1)$, where $f_1$ is the magnetic flux in the junction, normalized by the flux quantum, and $I_{c1}(0)$ is the critical current at zero field. So, the critical current of each junction should approach zero at $f=2$ since, geometrically, $f_1\approx f/2$. Therefore the resistance of each junction has a maximum at $f=2$ or a slightly higher field. This maximum negates the expected resistance minimum, related to the matching of the vortex lattice and the lattice of the Nb islands. To explain a zero or a very low critical current, we suggest that at $f=2$, the cores of the vortices enter the junctions (Fig.\ref{f1vortices}(b)). In such arrangements, the vortices are located at the maxima of their potential energy, so they can easily slide under the influence of the Lorentz force related to the bias current.

So far we have presented the multiple-island interference and the Fraunhofer diffraction effects, as they appear on the R-B curve. The same phenomena can been distinguished on the critical current versus magnetic field (Ic-B) curve (Fig.\ref{Ic_B}(top)). The most pronounced feature is the extra sharp peak at zero field. It reflects a coherent addition of the condensate transmission coefficients on all columns of the array. This is in analogy with the optical diffraction grating, which shows a high transmission at zero angle because the waves diffracted by all the slits arrive in phase. Such effect allows, in our case, very sensitive detection of the magnetic field, as illustrated in Fig.\ref{Sensitivity}. There, each point is obtained by measuring 1000 V-I curves, over a time span of 100 s, and subsequent averaging of the corresponding critical currents. The closest separation between the points we could resolve, using a simple dc-current measurement of the critical current, is about 8 nT. One key factor in such an approach is to realize a sample with a small fluctuation of the switching current. Our array shows a very low fluctuation, as is evident from Fig.\ref{Fig_RT-VI}(b, Insert), where each curve is measured four times. The standard deviation of the critical current was about 2.8 nA, which is about 0.1$\%$ of the critical current value. 

\begin{figure}[t!]
\centering
\includegraphics[width=9cm]{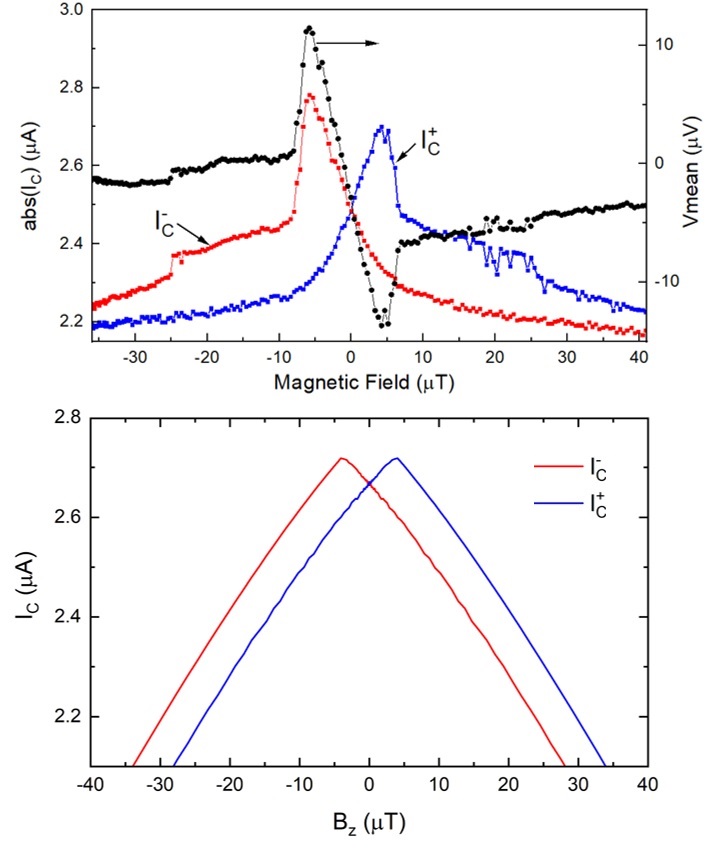} 
\caption{\label{Diode1}
\textbf{Rectifier effect}.  
Top panel: mean voltage, critical current on positive and negative branches plotted as a function of magnetic field.
Bottom panel: theoretical dependencies $I_C^+(B_z)$ (blue line) and $I_C^-(B_z)$ (red line), which are
obtained by numerical solution of Eqs. (\ref{IJ},\ref{phi1}) 
with $\tilde f=B_z(d+w)(d+0.3537w)/\phi_0$
and other parameters given in the caption of Fig. \ref{Ic_B}.}
\label{diode}

\end{figure}

As we zoom in on the central peak of the critical current (see Fig.\ref{diode}(top)), we find a shift of the point of the maximum away from the zero field point. Note that the point of zero field is defined through the equality of the two critical currents with the opposite polarity, i.e., $I_c^+=I_c^-$. The strongest asymmetry, which amounts to $\eta=I_c^-/I_c^+\approx1.2$, takes place at about -5 $\mu$T. At this point, $I_c^-=2.78\mu$A and $I_c^+=2.3\mu$A. Such 20$\%$ strong asymmetry of the critical current constitutes a field-controlled superconducting diode effect. It can be used to rectify ac-signals if the amplitude of the applied ac current is larger than the critical current of one polarity but lower than the the critical current of the opposite polarity. We demonstrate the diode effect directly by applying an ac current (10 Hz) with an amplitude that is lower than $\max(I_c^-)$ but larger than $\max(I_c^-)/\eta$. The result, a clear rectification effect, is shown in Fig.\ref{diode} by the black curve.

It might be instructive to compare these results to previously published observations of analogous phenomena. Recently, a superconducting diode was realized using a noncentrosymmetric superlattice. The lattice was made of alternating epitaxial films of
tantalum, vanadium, and niobium\cite{Ando}. In their study, the inversion symmetry was broken in the vertical direction. Thus, in order to break the time-reversal symmetry, a precisely aligned in-plane magnetic field, applied perpendicular to the current, was required. Our superconducting array, based on topological Bi$_{0.8}$Sb$_{1.2}$Te$_3$ films, requires much weaker fields in order to act as a superconducting rectifier. The magnetic field needs to be applied perpendicular to the plane of the sample but does not require any fine alignment.

{\bf Model}

We model the system as a two-dimensional array of wide Josephson junctions between square Nb islands
of the size $w\times w$ separated by the gaps of width $d$. The Josephson current flows
through the proximized BST layer, which has small induced superconducting gap $\Delta$ under the islands
and no gap between the islands. The
physics of Josephson arrays is very rich. For example, they exhibit
superconductor - insulator phase transitions\cite{Mooij,Haviland,Han,Marcus}, and 
at finite magnetic field a vortex lattice is formed in the array\cite{Lobb}. 
For these reasons, the full theoretical description of such systems is complicated\cite{Teitel,Halsey,Nori}. 
To keep the analysis tractable, 
here we adopt a simple model, which is formally valid at small magnetic fields. 
Despite its simplicity, the model captures the main effect we study --- Josephson current rectification. 

In the experiment we observe ac bias current rectification at small but finite magnetic fields, see Fig. \ref{diode}.
This effect may be caused by several reasons. First, it may be a simple self-field effect earlier observed in asymmetric SQUIDs\cite{Fulton,Gurtovoi,Murphy2, Dausy,Ilin}
and asymmetric wide junctions\cite{Monaco} with high critical current. Second, the rectification may be caused by
spin-orbit interaction in BST, which requires the presence of the two field components:
parallel ($B_y$) and perpendicular ($B_z$) to the BST plane. 
This effect has been previously demonstrated in junctions made on GaAs\cite{Baumgartner} and Bi$_2$Se$_3$\cite{Assouline} substrates.
We have verified that these two phenomena cannot explain our observations because 
both applied and induced fields are very small. Thus, our experiment
resembles recent studies of NbSe$_2$ nanowires\cite{Paradiso} and of thin metallic films\cite{Hou}, 
where the rectification has been observed in presence of only the $B_z$ field component. 
We attribute the rectification effect in our sample to the finite kinetic inductance of the proximized BST layer
under the Nb islands, as will be explained below. However, we cannot exclude an additional contribution from the edge states
in BST, which may effectively form asymmetric SQUID loops with a non-sinusoidal current-phase relation\cite{Chen}.

Let us consider the gap between two neighboring rows of Nb islands in the middle of the array, namely
the gap between the rows with the numbers $i$ and $i+1$. 
We express the Josephson current as an integral along the line 
going through the middle of this gap,
\begin{equation}
I_J = \int_{0}^{W} dy \, j_C(y)\sin[\phi(y)].
\label{IJ}
\end{equation} 
Here $W=N(w+d)$ is the total width of the array, $N$ is the number of the islands in one row ($N=23$ in our sample), 
$y$ is the coordinate along the line,
$j_C(y)$ is the coordinate dependent critical current density, and 
$\phi(y)$ is the gauge invariant phase difference across the Josephson junctions.
Here we assume the usual sinusoidal current-phase relation for the junctions.
The current density $j_C(y)$ equals to zero in the gaps between the Nb islands.
In Eq. (\ref{IJ}) the gauge invariant phase $\phi(y)$ is expressed as
\begin{equation}
\phi(y) = \varphi(y) - \frac{2\pi B_z d}{\phi_0}y,
\label{phi0}
\end{equation}
where $\varphi(y)=\varphi_{i+1}(y)-\varphi_i(y)$ is the difference between the phases of the superconducting order parameters 
in the two adjacent islands belonging to the rows $i+1$ and $i$, and $\phi_0$ is the magnetic flux quantum.

To elucidate the well-known relation between Eq. (\ref{IJ}) and the diffraction phenomenon in optics, we consider an 
ideal sample with constant current density $j_C(y)$ within the junctions and $j_C(y)=0$ in the gaps between them. 
We also assume that superconducting leads are bulky and the magnetic field is fully expelled from them.
In this case, the integral (\ref{IJ}) can be solved analytically. 
Maximizing the Josephson current over the coordinate independent phase difference $\varphi$,
one arrives at the familiar Fraunhofer-like dependence of the critical current on $f$,   
\begin{eqnarray}
I_C(f)=I_C^{(1)}(f)\left|\frac{\sin(\pi Nf)}{\sin(\pi f)}\right|,
\label{Fraunhofer}
\end{eqnarray} 
where   
\begin{eqnarray}
I_C^{(1)}(f)= \frac{I_C}{N}\left|\frac{w+d}{\pi f w}\sin\left(\frac{\pi f w}{w+d}\right)\right|
\label{IC1}
\end{eqnarray}
is the critical current versus flux dependence for a single junction
and $I_C$ is the critical current of the whole array at $f=0$. 
In Eqs. (\ref{Fraunhofer},\ref{IC1}) $f=B_z(w+d)d/\phi_0$, where $B_z$ is the field in the
gaps between the bulk islands. Due to flux focusing, this field is larger than the external field created by the magnet. Flux conservation ensures that
one can equivalently express the parameter $f$ by Eq. (\ref{f_def}), in which $B$ is the external field.
While this simple model well describes a single wide Josephson junction\cite{Barone}, 
it looses accuracy for a 2d array above certain value of the magnetic field where
vortex penetration becomes important.

To find the rectification effect, we start from the expression for the current density in the proximized BST layer
under a single square Nb island. Adopting the disordered superconductor model, we obtain 
${\bm j}=(\pi\Delta/2eR_\Box)(\nabla\varphi - 2e{\bm A}/\hbar c)$. Here $\varphi$ is the phase and
$\Delta$ is the absolute value of the induced
superconducting order parameter in BST and ${\bm A}$ is the vector potential with the components $A_x=B_zy$, $A_y=A_z=0$. 
We assume that the current flows along the $x$-axis, and we also fully ignore the screening of the field by thin Nb islands. 
Since $\nabla{\bm j}=0$ and 
$\nabla{\bm A}=0$, the phase under the island satisfies the equation $\nabla^2\varphi(x,y)=0$ with the boundary conditions 
derived from the conditions for the current components $j_x(0,y)=j_x(w,y)=j_C(y)\sin[\phi(y)]$, $j_y(x,0)=j_y(x,w)=0$.
Solving the equations for the phase $\varphi(x,y)$ under the two neighboring Nb islands from the rows $i$ and $i+1$
and taking the difference between the phases at the edges of these islands, 
we arrive at the self-consistent equation for the gauge invariant phase (\ref{phi0}) in the form
\begin{eqnarray}
\phi(y) &=& \varphi_0
-\int_0^w \frac{dy'}{2\pi}
\ln\frac{\left( \cosh\pi -\cos\frac{\pi(y-y')}{w} \right)\left( \cosh\pi -\cos\frac{\pi(y+y')}{w} \right)}
{\left( 1-\cos\frac{\pi(y-y')}{w} \right)\left( 1-\cos\frac{\pi(y+y')}{w} \right)}
\nonumber\\ &&\times\,
\left( \frac{2eR_\Box}{\pi\Delta} j_C(y')\sin[\phi(y')] + \frac{2\pi B_z}{\phi_0}y' \right)
-\frac{2\pi B_z d}{\phi_0}y.
\label{phi}
\end{eqnarray} 
Here $\varphi_0$ is an arbitrary constant phase shift.
Assuming further that the current density
$j_C(y')\sin[\phi(y')]$ slowly varies with the coordinate $y'$,
one can re-write Eq. (\ref{phi}) in the simplified form
\begin{eqnarray}
\phi(y) \approx \tilde \varphi_0  - \frac{2eR_\Box w}{\pi\Delta} j_C(y)\sin[\phi(y)] 
- \frac{2\pi \tilde f}{d+w} y.
\label{phi1}
\end{eqnarray} 
Here $\tilde\varphi_0$ is independent of the $y$ part of the phase and 
$
\tilde f={B_z(d+w)(d+0.3537w)}/{\phi_0}.
$
This parameter slightly differs from the normalized flux $f$ defined in Eq. (\ref{f_def}) due to field penetration in the leads. 
Eq. (\ref{phi1}) is valid in the limit $|\tilde f|\ll 1$, 
where one can assume that the current densities at both edges of a square island perpendicular to the bias current are equal,
and at the edges parallel to the bias current --- equal to zero. 
At higher fields local currents in the array flow in all directions and the field dependence of the critical
current should be determined by $f$, i.e. in the limit $|f|\gtrsim 1$ we expect $\tilde f = f$.
Eq. (\ref{phi1}) can be extended to the whole width of the array $0<y<W$, while Eq. (\ref{phi})
is valid only for a single junction, i.e. for $0<y<w$.
The second term in Eq. (\ref{phi1}) may be interpreted as the kinetic inductance contribution to the phase difference. 
It is equal to the phase drop across a single island. If this term is independent of $y$
it can be absorbed into the phase $\tilde\varphi_0$ and, therefore,  has no effect on the value of the critical current.
However, for the asymmetic current distribution $j_C(y)$ the kinetic term acts
as the effective self-induced field $B_z^{\rm eff}\propto j_C(y-W/2) - j_C(W/2-y)$. 
The effective field $B_z^{\rm eff}$ changes sign depending on the direction of the bias current and in this way causes the
rectification effect. Indeed, the maxima of the curves $I_C^{\pm}(\tilde f)$ shift from zero
in the opposite directions and occur at $\tilde f = f_0$ for $I_C^+$ and at $\tilde f = -f_0$ for $I_C^-$ with $f_0\propto B_z^{\rm eff}$.

Evaluating the Josephson current (\ref{IJ}) with $\phi(y)$ obtained from Eq. (\ref{phi}) requires extensive
numerical simulations. However, one can estimate the recitification effect on the basis of the simplified equation (\ref{phi1}). 
First, we consider the limit of small sheet resistance such that
\begin{eqnarray}
\alpha=\frac{2e I_C R_\Box}{\pi N\Delta}\ll 1.
\label{alpha}
\end{eqnarray} 
In this limit, one can derive a general expression for the shift of the critical
current maximum $f_0$ for an arbitrary critical current distribution $j_C(y)$,
\begin{eqnarray}
f_0 = \frac{e R_\Box w(w+d)}{\pi^2\Delta} \frac{\int_0^W dy_1 dy_2  j_C(y_1) j_C(y_2)  [j_C(y_1) - j_C(y_2)] (y_1-y_2)}
{\int_0^W dy_1  dy_2 j_C(y_1) j_C(y_2) (y_1-y_2)^2}.
\label{f0}
\end{eqnarray}
One can verify that $f_0\not=0$ only if the distribution $j_C(y)$ is asymmetric, i.e. $j_C(y-W/2) \not= j_C(W/2-y)$.
The critical current versus flux dependence in this approximation becomes
\begin{eqnarray}
I_C^\pm(\tilde f)=I_C - \frac{\pi^2(\tilde f\mp f_0)^2}{(d+w)^2I_C} \int_0^W dy_1\int_0^W dy_2\, j_C(y_1)j_C(y_2)(y_1-y_2)^2.
\label{ICpm}
\end{eqnarray}
For the simplest asymmetric critical current distribution of the form 
\begin{eqnarray}
j_C(y)=\frac{I_C}{W}\left[1+\beta \left(\frac{y}{W} - \frac{1}{2} \right)\right],
\label{asym}
\end{eqnarray}
where $\beta$ is a dimensionless asymmetry parameter, one finds 
\begin{eqnarray}
f_0 &=& \frac{eR_\Box  I_Cw\beta}{\pi^2 N^2\Delta(w+d)},
\label{f00}
\\
I_C^\pm(f) &=& I_C\left[1 - \frac{\pi^2 N^2}{6}\left(1 - \frac{\beta^2}{12}\right)(\tilde f\mp f_0)^2\right].
\label{ICpm1}
\end{eqnarray}
We note for clarity that the distribution (\ref{asym}) ignores the gaps between the islands because in our sample $d\ll w$.
Thus, in this approximation the array is essentially replaced by a single junction with the width $W$.
From Eq. (\ref{ICpm1}) one can estimate the asymmetry parameter introduced earlier,
\begin{eqnarray}
\eta=\frac{I_C^-(-f_0)}{I_C^+(-f_0)}=\left[1 - \frac{2}{3\pi^2}\left(\frac{eR_\Box I_Cw}{N\Delta(w+d)}\right)^2\beta^2\left(1 - \frac{\beta^2}{12}\right)\right]^{-1}.
\label{eta}
\end{eqnarray}

Let us check if Eq. (\ref{f00}) is consistent with the experiment.
From the experimental $I_C^\pm(B_z)$ curves presented in Fig. \ref{diode} we estimate $f_0\approx 4\times 10^{-3}$. 
Some of the experimental parameters are known from the sample design and from independent measurements, 
namely we can take $N=23$, $w=1.15$ $\mu$m, $d=150$ nm,
$I_{C}=2.7$ $\mu$A, $R_\Box=1.135$ k$\Omega$. The asymmetry parameter $\beta$ can be roughly estimated from the
spread of the re-trapping currents visible on the I-V curves presented in Fig.  \ref{Fig_RT-VI}b. 
Indeed, difference in the re-trapping currents of individual junctions 
should roughly correspond to the difference in their critical currents.
Assuming further that the junctions with the highest and the lowest critical currents are located
close to the opposite edges of the array, one can roughly approximate the distribution $j_C(y)$ by Eq. (\ref{asym}) 
with the asymmetry parameter $\beta\approx 0.15$.
With these parameters, Eq. (\ref{f00}) gives the experimentally observed value $f_0\approx 4\times 10^{-3}$
if one chooses the gap in the BST layer as $\Delta=19$ $\mu$eV. This value is approximately 18 times smaller than peviously reported one
in similar structures\cite{Tikhonov}, but it is consistent with the data from angle-resolved photoemission spectroscopy,
which show strong suppression of the proximty effect in the BST\cite{Hlevyack}.
Moreover, the value of the gap is consistent with the $I_C^{\rm jct} R_{\rm norm}^{\rm jct}$-product for a single junction, 
where $I_C^{\rm jct}=I_C/N$ is the critical current and $R_{\rm norm}^{\rm jct}$ is the normal state resistance of the junction.
Indeed, $I_C^{\rm jct}=I_C/N=0.117$ $\mu$A and
the resistance of a single junction should be equal to the resistance of the whole square array at high temperatures, where Nb is superconducting but the proximity effect is negligibly weak (at about 6 K).
In this way, from Fig. \ref{RB} we estimate $R_{\rm norm}^{\rm jct}\approx 225\,\Omega$. 
With these parameters we obtain $I_C^{\rm jct} R_{\rm norm}^{\rm jct}\approx 26.4$ $\mu$V, which is indeed
close to the gap value $\Delta=19$ $\mu$eV obtained above. 
With these numbers the asymmetry parameter (\ref{eta})
takes the value $\eta=1.062$, which is a bit lower than the observed one, $\eta=1.2$.

Although the simple equations (\ref{f00},\ref{ICpm1}) explain the data reasonably well, they
have been derived under the assumption $\alpha\ll 1$. However, with the parameters outlined above one finds $\alpha \approx 4$. 
For this reason, we have lifted the restrictions on $\alpha$ and have solved Eq. (\ref{phi1}) numerically
to find the coordicate dependence of the phase $\phi(y)$. 
For $\alpha > 1$ this equation has multiple solutions with possible jumps between them. We choose one of these solutions
for every value of $y$, with phase jumps to another solution happening at certain values of this coordinate 
where the previous solution ceases to exist.
Such jumps correspond to the centers of the Josephson vortex cores, which are formed in the array. 
In addition, we assume constant critical current density, $j_{C,n}$, within the $n-$th junction, 
and assume that these values follow Eq. (\ref{asym}),
\begin{eqnarray}
j_{C,n}=\frac{I_C}{W}\left[ 1 + \beta\left(\frac{n-1}{N}-\frac{1}{2}\right) \right].
\label{jCn}
\end{eqnarray}
To get better agreement with the experiment, we have slightly adjusted the critical current of the whole array choosing $I_C=2.88$ $\mu$A, and have also changed the value of the gap taking $\Delta=59$ $\mu$eV.
In Fig. \ref{Ic_B} we plot the dependence $I_C^+(\tilde f)$, numerically obtained from Eqs. (\ref{IJ},\ref{phi1}), for the wide range of the
normalized fluxes $|\tilde f|\leq 2.5$. We observe qualitative agreement between the theory and the experiment
in spite of the simplicity of the model, which does not properly describe 
the complicated process of the vortex formation. 
In particular, both in the theory and in the experiment the peaks occurring at $\tilde f=1/2$, $\tilde f=1/3$ and $\tilde f=2/3$ are clearly visible. 

Moreover, the experiment shows a clear jump at $B=25\mu$T, which can be due to a vortex entrance, the phenomenon also found within the model. This might be not the first vortex entering. The first vortex entrance seems to be associated with a massive and and sharp drop of the critical current observed at about $B=6\mu$T. This magnetic field is close to the estimated value. Indeed, since the field required to populate each unit cell with exactly one vortex is 1.23 mT and the number of unit cells is 23x23=529, the field needed to put just one vortex in the array is 2.3$\mu$T. The experimental field at which a vortex can enter ($6\mu$T) is somewhat larger, probably due to some nonnegligible screening.

If one further examines Fig. \ref{Ic_B}, one finds an interesting distinction between the experiment (top) and the model (bottom). Namely, the model predicts that at various values of the magnetic flux the critical current should approach zero. Yet the experiment shows that in the entire range of fields the critical current is never zero. This could be due to some inhomogeneity that is qualitatively different from the one included in the model. Alternatively, this could be an instance of a node lifting, which might indicate that vortices carry Majorana zero modes that contribute the measured critical current\cite{VanHarlingen1}. A qualitatively different experiment will be needed to confirm or reject this interesting possibility. For example, applications of microwave pulses may be able to push vortices around and entangle them, thus producing braiding and the corresponding modifications of the critical current. This will be the subject of the future work.

To reveal the rectification effect,  
in Fig. \ref{diode} we consider small magnetic fields, $|B_z|\leq 40$ $\mu$T, which correspond to $|\tilde f|\leq 0.0113$.
The maxima of the theoretical curves $I_C^+(B_z)$ and $I_C^-(B_z)$ occur at $B_z=\pm 4$ $\mu$T, which is close
to the experimental values $\pm 5$ $\mu$T. The asymmetry parameter estimated from the theoretical curves is again lower than 
in the experiment, $\eta=1.045$.
The theoretical curves differ from the experimental ones in shape.
To reproduce the experimental $I_C^\pm(B_z)$ curves more accurately one should precisely know the distribution of the critical current
densities $j_{C,n}$. Considering these uncertainties, we conclude that our model reasonably well describes the experimental data.

{\bf Conclusions} 

We observe both the interference and the diffraction effects on a superconducting array involving an intrinsic topological insulator. The system is analogous to an optical diffraction grating. The interference, originating from multiple superconducting islands, produces a very sharp peak of the critical current. The diffraction effect causes the second matching peak to disappear. We develop a model which captures the main features of the experiment. One aspect in which the model differs from the experiment is that the model predicts that at many various values of the magnetic flux the critical current should return to zero due to either interference or the diffraction. Yet the experiment demonstrates that the critical current never approaches zero. Such so-called node lifting could be a sign that a more advanced model of the disorder is needed. It is also consistent with a more exotic phenomenon of node lifting caused by Majorana zero modes\cite{VanHarlingen1}. More advanced experiments will be needed to distinguish these two possibilities.

The sample also exhibits a superconducting diode effect. Our model explains it based on the assumption of a significant kinetic inductance of the topological surface states combined with some inhomogeneity. The rectification appears at very lower fields of the order of a few micro-Tesla. Finally, we find that the devices can be used as sensitive absolute magnetic field detectors due to the interference phenomenon analogous to an optical diffraction grating.

{\bf Methods} 

The substrate for TI growth is a single-side polished c-plane sapphire (MTI Corp) that was annealed in oxygen at 1100 $ ^\circ C $ for several hours for atomic-level flatness and large crystalline terraces. The unpolished side was coated with 100 nm Titanium to allow for accurate measurements of substrate temperature using a pyrometer, as sapphire is transparent in the infrared spectrum. The substrate was clamped onto a tantalum holder using tantalum wire at its four corners and allowed to outgas in the load lock for several hours before being introduced into the growth chamber for high-temperature outgassing. 
The growth chamber base pressure was at a low $10^{-8}$ torr level. 

The composition of the film was controlled by the flux ratio of Bismuth/Antimony from Knudsen effusion cells with an excess Tellurium flux to compensate for tellurium losses from the surface. A Te/(Bi+Sb) ratio greater than 6 is sufficient. Two quartz crystal microbalance (QCM) measurements of the Bi and Sb fluxes were taken to ensure the fluxes are stable. The substrate was heated to the growth temperature of 200-215 $ ^\circ C $ on the pyrometer (260$ ^\circ C $ on a thermal couple placed on the backside of the substrate) and the tellurium shutter was opened for 10 seconds to ensure the starting layer of the quintuple stacked  Te-(Bi+Sb)-Te-(Bi+Sb)-Te structure is tellurium. Then the Bi and Sb shutters were opened and shut simultaneously for periods of time calculated using their flux measurements. A two-stage growth method was used, where the first 3 quintuple layers (QLs) are grown and annealed in an excess of Te flux for an hour before growing the remaining 37 QLs. At the end of film growth, the film was annealed for 4 hours at around 260-265$ ^\circ C $ on the pyrometer (330$ ^\circ C $ on the thermal couple), but with a lowered Te flux to prevent excess Te buildup. Once the annealing was completed, the tellurium flux was cut off. The film was then moved into the load lock and allowed to fully cool down, before being removed from the vacuum. To prevent oxidation from happening at the exposed film surface due to exposure to the atmosphere, polymethyl methacrylate (PMMA) was dropped onto the film immediately when it was removed and allowed to dry. PMMA layer was completely removed prior to device fabrication by dissolving the protective PMMA film in acetone and rinsing the sample in isopropyl alcohol.

 For device fabrication, a two-session electron beam (e-beam) lithography method was used. In the first session the film was coated with PMMA 950 A4 and exposed under a 10kV e-beam to define the array structure,  electrodes and alignment markers. The sample was developed in Methyl isobutyl ketone (MIBK):IPA 1:3 solution for 25 seconds and immediately loaded into a plasma sputtering system for metal deposition. A two-second ion milling was applied to briefly clean the BST surface, followed by 30 nm Nb sputtering. Lift off was done by placing the sample inside acetone at room temprature for 2 hours and sonicating it for 5 seconds. In the second lithography session the trenches are aligned and exposed on the sample. After developing in MIBK:IPA 1:3 solution, a 60 second ion milling session was applied to completely remove the TI, such that the electrodes for current and voltage probing are defined and separate from each other.

\hskip 1pt

\begin{addendum}
\item The work was supported in part by the NSF DMR-2104757 and  by the NSF OMA-2016136 Quantum Leap Institute for Hybrid Quantum Architectures and Networks (HQAN). 
This project has also received funding from the European Union’s Horizon 2020 research and innovation program under Grant Agreement No. 862660/QUANTUM E-LEAPS. 
We are grateful to D. J. Van Harlingen for useful discussions.

\item [Author contribution] 
The work was conceived and planned by X.S., J.N.E., and A.B. High quality topological insulator films have been grown by S.S-B, Y.B. and J.N.E. The experiments and data analysis have been performed by X.S., I.B., A.R., and E.I. The theoretical analysis was developed by D.G. All work was carried out under the supervision of J.N.E. and A.B.

\item[Competing interests] The authors declare that they have no competing financial interests.

\item[Correspondence] Correspondence and requests for materials should be addressed to bezryadi@illinois.edu.

\end{addendum}

\end{document}